\documentclass[pss]{wiley2sp} 
\usepackage{amsmath}
\usepackage{bm}             

\begin{document}

\title{Ga$_{x}$Cu$_{4-x}$(OD)$_{6}$Cl$_{2}$: Insulating ground state in an electron doped kagome system}

\author{%
  P.~Puphal\textsuperscript{\Ast,\textsf{\bfseries 1,2}},
  K.~M. Ranjith\textsuperscript{\textsf{\bfseries 3}},
  A.~Pustogow\textsuperscript{\textsf{\bfseries 4}},
  M.~M\"uller\textsuperscript{\textsf{\bfseries 1}},  
  A.~Rogalev\textsuperscript{\textsf{\bfseries 5}},
  K.~Kummer\textsuperscript{\textsf{\bfseries 5}},
  J.-C.~Orain\textsuperscript{\textsf{\bfseries 6}}, 
  C.~Baines\textsuperscript{\textsf{\bfseries 6}},  	
  M.~Baenitz\textsuperscript{\textsf{\bfseries 3}},
  M.~Dressel\textsuperscript{\textsf{\bfseries 4}},
  E.~Kermarrec\textsuperscript{\textsf{\bfseries 7}},
  F.~Bert\textsuperscript{\textsf{\bfseries 7}},
  P.~Mendels\textsuperscript{\textsf{\bfseries 7}},
  C.~Krellner\textsuperscript{\Ast,\textsf{\bfseries 1}}
}

\mail{\textsf{pascal.puphal@psi.ch; krellner@physik.uni-frankfurt.de}}

\institute{%
  \textsuperscript{1}\,Physikalisches Institut, Goethe-Universit\"at Frankfurt, 60438 Frankfurt/Main, Germany\\
  \textsuperscript{2}\,Laboratory for Multiscale Materials Experiments, Paul Scherrer Institute, 5232 Villigen, Switzerland\\
  \textsuperscript{3}\,Max Planck Institute for Chemical Physics of Solids, 01187 Dresden, Germany\\
  \textsuperscript{4}\,1.~Physikalisches Institut, Universit\"at Stuttgart, 70569 Stuttgart, Germany\\
  \textsuperscript{5}\,Electronic Structure, Magnetism \& Dynamics Group, ESRF, 38043 Grenoble, France\\ 
     \textsuperscript{6}\,Laboratory for Muon Spin Spectroscopy, Paul Scherrer Institute, 5232 Villigen PSI, Switzerland\\
   \textsuperscript{7}\,Laboratoire de Physique des Solides, Universit\'e Paris-Sud, 91405 Orsay, France\\
}


\abstract{\bf%
We present a successful synthesis method of the series Ga$_{\bm{x}}$Cu$_{\bm{4-x}}$(OD)$_{\bm{6}}$Cl$_{\bm{2}}$ with substitutions up to $\bm{x=0.8}$. The compound remains a frustrated kagome system with an insulating ground state similar to herbertsmithite for these substitutions, as the additional charge of Ga$^{3+}$ is most likely bound to additional OH$^{\bm{-}}$ and Cl$^{\bm{-}}$. Besides infrared measurements, we present magnetic and specific-heat data down to 2~K for selected samples with $\bm{0\leq x\leq0.8}$. With increasing $\bm{x}$ the long-range magnetic order is suppressed, similar to what was observed in the series Zn$_{\bm{x}}$Cu$_{\bm{4-x}}$(OH)$_{\bm{6}}$Cl$_{\bm{2}}$, indicating that Ga goes predominantly to the inter-plane position of the layered crystal structure. The reduction of the frozen magnetic fraction with increasing substitution was followed by $\bm{\mu}$SR measurements. $\bm{^{69,71}}$Ga nuclear magnetic resonance (NMR) was applied as a local probe for Ga induced disorder. One well resolved Ga NMR line of moderate width is found for each isotope across the phase diagram which indicates a rather homogeneous distribution of the Ga isotopes on a single site in the lattice.}

\maketitle   

\section{Introduction}
The perspective of a quantum spin-liquid state created a large interest in geometrically frustrated materials. Realized at low temperatures it is a highly correlated state that has no static long-range magnetic order, despite sizable magnetic interactions \cite{Balents2010}. Compounds with decoupled antiferromagnetic kagome layers are one experimental setting to search for a realization of the quantum spin-liquid state, where herbertsmithite, ZnCu$_{3}$(OH)$_{6}$Cl$_{2}$, has become one of the most prominent materials in recent years \cite{Mendels2007,Helton2007,Han2012,Shores2005}.

The dominant magnetic interaction in herbertsmithite is caused by Cu-O-Cu antiferromagnetic superexchange with an exchange energy of J $\sim17$~meV, but no magnetic long-range order has been observed down to 50~mK \cite{Helton2007}. Therefore, the spin-liquid ground-state of this material was investigated in great detail \cite{Norman2016}. However, a strong structural drawback of herbertsmithite is the intrinsic Zn-Cu-antisite disorder of up to 15\% \cite{Vries2012,Han2016}. The precise amount of this antisite disorder is difficult to quantify with conventional x-ray scattering techniques, causing some debate in the literature about the precise stoichiometry of these materials \cite{Norman2016,Freedman2010}. 

Recently, a high potential is seen in chemical doping a kagome system since Mazin et al. \cite{Mazin2014} have proposed that a correlated Dirac metal can be found in electron-doped herbertsmithite which might be realized by replacing Zn by a trivalent ion \cite{Mazin2014,Guterding2016}. There, GaCu$_{3}$(OH)$_{6}$Cl$_{2}$ was suggested as the ideal candidate for an electron doped kagome system, since even when including correlation effects to the calculation, the proposed bandstructure shows the exotic phenomena of a Dirac metal. Experimentally, it turned out to be challenging to synthesize samples of doped herbertsmithite. When replacing divalent Zn by trivalent Y, Eu, Nd etc. the additional electron is either bound to additional OH$^{-}$ or Cl$^{-}$ anions and the kapellasite type structure P$\bar{3}$m1 is realized \cite{Sun2016,Puphal2017,Puphal2018}. A direct doping attempt by intercalating Li into herbertsmithite also does not lead to metallic behavior but instead strong Anderson localization of the additional charge carriers was reported \cite{Kelly2016}. 

In this paper, we report on the successful synthesis of Ga-substituted atacamite Cu$_{4}$(OD)$_{6}$Cl$_{2}$, which is the first report realizing the herbertsmithite type R$\bar{3}$m1 structure containing a trivalent ion on the Cu site. We analyze the magnetic behavior down to 2 K and found similar characteristics as in the series Zn$_{x}$Cu$_{4-x}$(OH)$_{6}$Cl$_{2}$. We have applied bulk methods including specific heat and magnetization, optical and x-ray absorption spectroscopy (XAS), as well as local probes like muon spin relaxation ($\mu$SR) and nuclear magnetic resonance (NMR) to investigate in detail the magnetism of this series. Especially the muon relaxation could probe tiny internal field effects whereas the NMR is rather sensitive for local disorder and the NMR line shift corresponds to the local susceptibility. In contrast to Zn$_{x}$Cu$_{4-x}$(OH)$_{6}$Cl$_{2}$ where NMR was performed on $^{17}$O and $^{35,37}$Cl, Ga-substituted atacamite opens up the opportunity to perform $^{69,71}$Ga NMR as a probe for local disorder and magnetism. In contrast to Zn, the two natural isotopes of Ga with $S=3/2$ possess a rather large nuclear moment suitable for NMR, with a total natural abundance of 100 \% for the two isotopes. 

\section{Experimental Details}
The energy-dispersive x-ray (EDX) spectra were recorded with an AMETEK EDAX Quanta 400 detector in a Zeiss DSM 940A scanning electron microscope (SEM). Powder x-ray diffraction (PXRD) were collected on a Bruker D8 Focus using a Cu x-ray generator and the Rietveld refinement was done using the fullprof suite \cite{Rodriguez-Carvajal1993}.

X-ray absorption spectroscopy (XAS) at the Cu and Ga \textit{K} edges has been performed at the ID12 beamline of the ESRF. For the measurements the powders were pressed to a pellet. The spectra were obtained at room temperature by recording the total fluorescence yield from the samples as a function of incident photon energy using a photodiode mounted at 90 degrees with respect to the incident beam. The signal of a 4 $\mu$m thin Ti foil mounted upstream the samples was used to normalise the data to the incident photon flux. As reference samples we used powder samples from Sigma Aldrich (CuO, Cu$_{2}$O, Ga$_{2}$O$_{3}$), a Cu metal foil as well as single crystals of Ga$_{3}$Ga$_{5}$O$_{12}$ and GaN. The reference spectra were recorded immediately after the sample spectra using the same setup and the same experimental conditions.

Optical characterization in the range 20--20000~cm$^{-1}$ was performed with a Bruker Fourier-transform infrared spectrometer. We pressed pellets of various thicknesses for our reflection ($d\approx 10^{-3}$~m) and transmission ($d\approx 10^{-5}$--$10^{-4}$~m) measurements. While $\sigma_1(T,\omega)$ was obtained by fitting the Fabry-Perot fringes in the THz range (20--50~cm$^{-1}$), the far-infrared (50--700~cm$^{-1}$) optical conductivity was determined using the Kramers-Kronig relations, in agreement with the direct calculation from $R$ and $T$.

The specific-heat data and magnetic measurements were collected with the standard options of a Physical Property Measurement System from Quantum Design in a temperature range of 1.8 to 300~K. $\mu$SR measurements were performed in zero field (ZF), longitudinal applied field (LF) or weak transverse field (WTF), with respect to the muon initial polarization at both GPS and LTF at PSI, giving a combined temperature regime extending from 50~mK up to 50~K. 

NMR experiments were performed using phase-coherent pulsed Tecmag-Apollo NMR spectrometer. $^{69,71}$Ga NMR spectra were measured using a field sweep NMR point-by-point spin-echo technique. The NMR field sweep spectra are obtained by integration over the spin echo in the time domain at a given magnetic field. The magnetic field was calibrated by using a non magnetic reference compound (GaAs single crystal). As $^{69}$Ga and $^{71}$Ga are twin isotopes the two lines should have an intensity ratio of $^{69}I/^{71}I = 1.5$ (natural abundance of 60.4\% for $^{69}$Ga and 39.6\% for $^{71}$Ga) and the ratio of the resonance fields should correspond to the ratio of the gyromagnetic ratios ($^{69}$($\gamma$/$2\pi$)\,=\,10.219 MHz/T and $^{71}(\gamma$/$2\pi)=12.983$~MHz/T) which leads to $^{69}\gamma/^{71}\gamma= 1.270$. Samples of Ga-substituted atacamite are deuterated (nearly 100\% replacement of $^{1}$H) which in addition allows for deuterium $^{2}$H NMR (with spin $I=1$).

\section{Synthesis}

We managed to synthesize powder samples of Ga$_{x}$Cu$_{4-x}$(OH)$_{6}$Cl$_{2}$, with its green color in contrast to the assumption of a Dirac metal. From a detailed study we found that the phase formation of Ga$_{x}$Cu$_{4-x}$(OH)$_{6}$Cl$_{2}$ happens only at low temperatures with the necessity of a surface to react on. For the synthesis we slowly dissolve CuO in a GaCl$_{3}$-H$_{2}$O solution of 0.1 to 1.1 M at 90$^{\circ}$C using the reflux method. For the reflux method in kapellasite \cite{Colman2008} glass beads were used to initiate the formation of the material. For Ga$_{x}$Cu$_{4-x}$(OH)$_{6}$Cl$_{2}$ this did not improve the synthesis, but it was found that Ga$_{x}$Cu$_{4-x}$(OH)$_{6}$Cl$_{2}$ forms preferably on the rough surface of unreacted CuO. 

GaCl$_{3}$ is strongly hygroscopic, so that it turns liquous at air and the reaction with water is extremely exothermal. The liquid itself is a Lewis acid and thus the solutions already heat up themselves when being mixed with water and for a controlled growth have to be cooled down again before the CuO is added. Since GaCl$_{3}$ is a strong Lewis acid with pH values of 1-3 in our molarity regime it already dissolves some CuO at room temperature. This leads to the formation of CuCl$_{2}$ and a coloration, in contrast to ZnCl$_{2}$ where the dissolved CuO directly forms herbertsmithite and the solution stays clear. Thus a higher amount of CuO has to be used during the synthesis since otherwise the whole CuO will be dissolved. Furthermore, we need additional CuO as a reaction-surface. A good ratio was found with (1.76\,g GaCl$_{3}$)/($\sim1$\,ml GaCl$_{3}\cdot x$H$_{2}$O) in 10 ml distilled water and 1.2\,g CuO for 1.0\,M = mol/l. The synthesis shows a good reproductivity and each molar ratio has been synthesized several times. 

\subsection{Deuteration}\label{Sec:Deu}
The same synthesis procedure was repeated with dehydrated GaCl$_{3}$ opened in a Desiccator on D$_{2}$O atmosphere, which leads to a liquid of GaCl$_{3}-x$ D$_{2}$O. Then the reflux method at 90$^{\circ}$C was applied to 0.72\,g CuO in a 6\,ml D$_{2}$O - GaCl$_{3}$ solution of 0.1 to 1.1 M. The resulting powder samples of Ga$_{x}$Cu$_{4-x}$(OD)$_{6}$Cl$_{2}$ have the same substitution amounts for given molar ratios and show the same physical properties. Here, we present the structural and magnetic properties for 6 different Ga concentrations listed in Table~\ref{tab:Average-results-of}. For comparison, a clinoatacamite sample ($x=0$) was prepared in a similar way using CuCl$_{2}-4$D$_{2}$O.

\begin{table}[h]
\caption{\label{tab:Average-results-of}Average substitution factor $x$ from EDX measurements on the Ga$_{x}$Cu$_{4-x}$(OD)$_{6}$Cl$_{2}$ powder series for the different batches as explained in Sec.\ref{Sec:Deu}. The given errors are not only instrumental but mainly due to the statistical distribution over several samples.}
\noindent \begin{centering}
\begin{tabular}{|c|c|c|c|c|c|c|}
\hline 
M& 0.1M & 0.2M & 0.4M & 0.7M & 1.0M & 1.1M \tabularnewline
\hline 
\hline 
$x$ & 0.09(6) & 0.19(4) & 0.4(1) & 0.47(8) & 0.8(1) & 0.7(2)\tabularnewline 
\hline 
\end{tabular}
\par\end{centering}
\end{table}

\subsection{CuO impurities}
As the compound forms on the surface of CuO and an excess due to the dissolution already at room temperature is necessary, there is always some residual CuO remaining after the synthesis. The amount of the CuO impurity phase is higher in low molarity solutions going up to 74 wt\% in 0.1 M. This impurity amount could however be reduced by sedimentation: The powder was shaken in distilled water and given a short time to let the CuO with a higher density (6.31\,g/cm$^{3}$ compared to 3.85\,g/cm$^{3}$ of Ga$_{x}$Cu$_{4-x}$(OD)$_{6}$Cl$_{2}$) fall down. Then the solution was decanted and filtered without the bottom part of CuO, successfully reducing the CuO amount to $\sim$7~wt\% for all deuterated batches except the 0.1 M since here the yield was very small after separation. Thus the sample was not measured in $\mu$SR as the impurity amount is too high. We note that the Ga$_{x}$Cu$_{4-x}$(OD)$_{6}$Cl$_{2}$ powder is really microcrystalline compared to the larger CuO crystallites and there is always some part lost during the sedimentation process reducing the yield enormously. In the case of the 0.1~M sample the segregation gave no yield and thus some of the original CuO containing powder was used. We determined the amount of CuO impurity in each batch by refining the PXRD patterns with the two phase parts and this was taken into account for the magnetization measurement by dividing by the real mass part of the Ga$_{x}$Cu$_{4-x}$(OD)$_{6}$Cl$_{2}$ phase.

\subsection{Additional synthesis attempts}
We were not able to obtain single crystals, because higher temperatures and higher molar ratios lead to decomposition of the Ga$_{x}$Cu$_{4-x}$(OH)$_{6}$Cl$_{2}$ samples. We could narrow down the decomposition of Ga$_{x}$Cu$_{4-x}$(OH)$_{6}$Cl$_{2}$ and thus formation temperature of GaO(OH) to temperatures above 100$^{\circ}$C. But all direct crystal growth attempts with a gradient similar to herbertsmithite gave pure clinoatacamite Cu$_{2}$(OH)$_{3}$Cl crystals in a GaO(OH) matrix. In case of higher molar ratios of GaCl$_{3}$ similar problems occur since all CuO is dissolved and GaO(OH) begins to fall out. Attempts of solution growth, where CuO was intentionally dissolved in a GaCl$_{3}$- H$_{2}$O solution only gave CuCl$_{2}$ single crystals in a green viscous liquid. The utilisation of Cu$_{2}$(OH)$_{3}$CO$_{3}$ to increase the pH value leads to the formation of a highly viscous liquid preventing any synthesis attempts. Experiments similar to \cite{Sun2016} with LiCl and LiOH flux yielded small single crystals of Cu$_{2}$(OH)$_{3}$Cl. Using Ga$_{2}$O$_{3}$ or presintered Ga$_{2}$CuO$_{4}$ and CuCl$_{2}$ also only formed clinoatacamite and Ga-/Cu oxides. Thus Ga$_{x}$Cu$_{4-x}$(OH)$_{6}$Cl$_{2}$ gives a challenging phase similar to kapellasite, which can so far only be obtained from the reflux method at low temperatures forming only as microcrystals.

\section{Structure}
Unlike other trivalent ions, which all realize the kapellasite type structure \cite{Sun2016,Puphal2017,Puphal2018} the compounds Ga$_{x}$Cu$_{4-x}$(OD)$_{6}$Cl$_{2}$ show similar PXRD patterns as for Zn substitutions. Thus Ga$_{x}$Cu$_{4-x}$(OD)$_{6}$Cl$_{2}$ has a similar structure as clinoatacamite and herbertsmithite. The diffraction reflexes show a slight broadening as it was observed for Zn$_{x}$Cu$_{4-x}$(OH)$_{6}$Cl$_{2}$ making it hard to differentiate clinoatacamite from herbertsmithite \cite{Freedman2010}. Since the electron density of Ga and Cu is quite comparable, PXRD will not be sufficient to differentiate those two atoms. Therefore, all diffraction patterns look rather similar, however the amount of CuO and other impurity phases varies. We performed a preliminary refinement of a PXRD for the 0.7 M batch. The goodness of the fit with the herbertsmithite structure R$\bar{3}$m is $\chi^{2}=3.888$ with a unit cell of $a=b=6.83463\thinspace\text{\AA}$, $c=13.95853\thinspace\text{\AA}$. In comparison to herbertsmithite ($a=b=6.8326(3)\thinspace\text{\AA}$ and $c=14.0686(8)\thinspace\text{\AA}$) the lattice parameters show mainly a smaller c-axis, in accordance with a smaller ionic radius of Ga compared to Zn. Since the c-axis is influenced most, it is plausible to assume that the Ga ion mainly goes into the Cu-position between the kagome planes. Furthermore, also theoretical calculations suggest no Ga on the intraplane Cu-positions \cite{Guterding2016}. A real refinement of the data with the Ga occupation on Cu sites as a free parameter cannot be performed, due to the similar scattering cross-section of these two atoms, a similar problem as it occurred for Zn and Cu in herbertsmithite. 

\begin{figure}[t]
\noindent \begin{centering}
\includegraphics[width=\columnwidth]{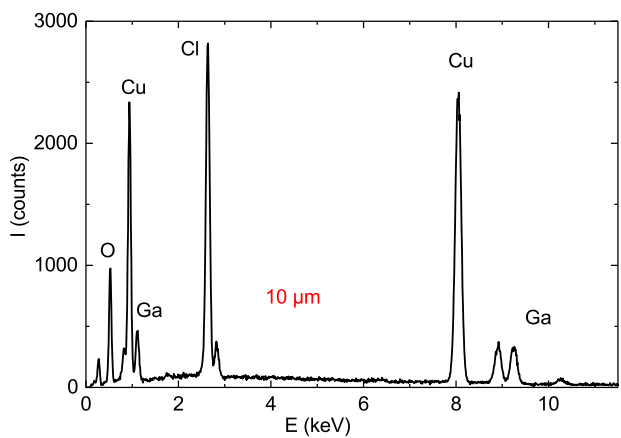}
\par\end{centering}
\caption{\label{EDX Ga}EDX analysis of a Ga$_{x}$Cu$_{4-x}$(OH)$_{6}$Cl$_{2}$ sample part from the 1.0~M solution. The inset image shows a SEM picture, where the red cross depicts the measured area. The brighter areas are due to GaO(OH) impurities.}
\end{figure}

\section{EDX Analysis}\label{Sec:EDX}
Since we obtained only powder samples of Ga$_{x}$Cu$_{4-x}$(OD)$_{6}$Cl$_{2}$, the EDX analysis was performed on pressed pellets. In Figure~\ref{EDX Ga} an exemplary EDX spectra is shown with the SEM image on the inset for a sample from the 1.0~M batch. The samples were coated by a carbon layer as the conductance was too low leading to charging effects, already indicating the insulating behavior of the sample. We note that the EDX analysis on powder samples with impurity phases causes some uncertainties for the measured elements. It is used as a general guide to have some insight on the substitution amount, since a refinement on a PXRD pattern of occupancies between neighboring ions such as Ga and Cu gives no reliable values. We determined the average value including its resulting statistical error for oxygen and chlorine over all measured points of all substituted batches with a minimum of four points per batch as an indication for the stoichiometry. The resulting value of $x$ is shown in Table~\ref{tab:Average-results-of}. We observed that within one batch the powder shows different substitution amounts with a difference up to 2 at\%, which is taken into account in the large statistical errors in Table~\ref{tab:Average-results-of}, which is further enhanced by amounts of GaO(OH) impurities (white parts in Figure~\ref{EDX Ga}).

For the example shown in Figure~\ref{EDX Ga}, the resulting at\% of the elements are 6.5(5) Ga, 23(1) Cu, 53(4) O, 18(1) Cl, which in this case would correspond to $x=0.9$. The values of oxygen and chlorine show always a slight enhancement, compared to the ideal herbertsmithite-type stoichiometry, where one would expect 50 at\% O and 16.7 at\% Cl, however, the accuracy of our measurements is not sufficient to finally prove that. However, as discussed in Sec.~\ref{SecDis} the most likely scenario is, that the additional charge of Ga is balanced by this additional Cl and OH, leading to a possible formula Ga$_{x}$Cu$_{4-x}$(OD)$_{6+2x/3}$Cl$_{2+x/3}$. For $x=0.9$, this would correspond to 6.9 at\%\,Ga, 24 at\%\,Cu, 51 at\%\,O, 17.8 at\%\,Cl which is within the observed values. Generally we see in EDX measurements that with increasing GaCl$_{3}$ content in the solution (molar ratio, first line of Table~\ref{tab:Average-results-of}), also the Ga content in the powder compound increases. But even at the highest possible molar ratios, which still lead to the proper phase, the average Ga content stays below $x=1$ (see Table~\ref{tab:Average-results-of}), which could be ascribed to the fact that the herbertsmithite-type structure is not stable with a full additional OH/Cl$^{-}$ molecule/atom. This is further being supported by the other cases with a trivalent atom on the Cu site, which realize the kapellasite type P$\bar{3}$m1 structure at only and exactly $x=1$ \cite{Sun2016,Puphal2017,Puphal2018}.

\begin{figure}[t]
\noindent \begin{centering}\includegraphics[width=\columnwidth]{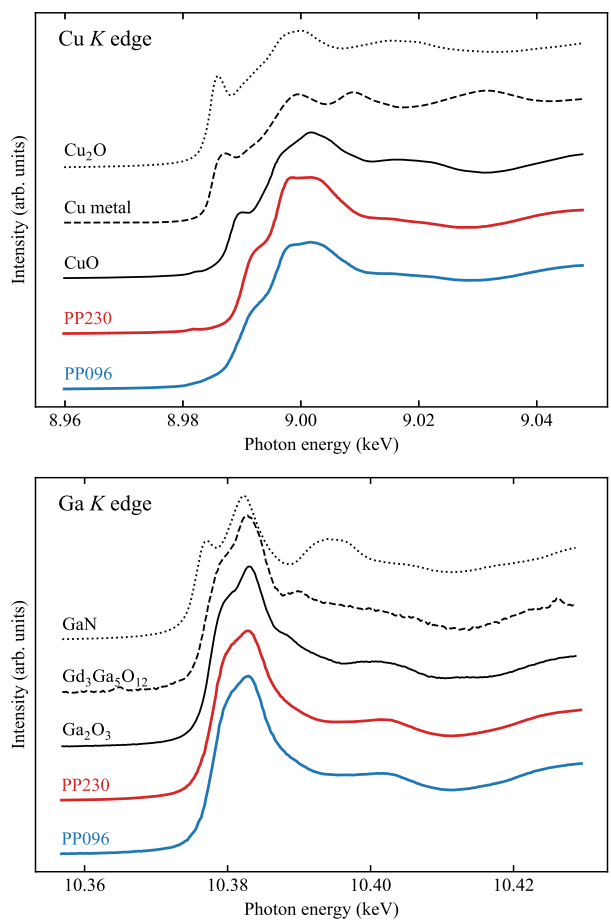}
\par\end{centering}
\caption{\label{XAS} XAS spectra of a 0.8 M (blue line) and 1.1 M (red line) sample measured with a 300 x 300 micrometer beam at 300\,K, together with the appropriate reference spectra.}
\end{figure}

\def\Herb{$\rm ZnCu_3(OH)_6Cl_2$}\def\GA{$\rm Ga_xCu_{4-x}(OH)_6Cl_2$}\def\GAD{$\rm Ga_xCu_{4-x}(OD)_6Cl_2$}\def\Y{$\rm Y_3Cu_9(OH)_{19}Cl_8$}\def\cm{cm$^{-1}$}

\begin{figure*}[t]
\centering
\includegraphics[width=2\columnwidth]{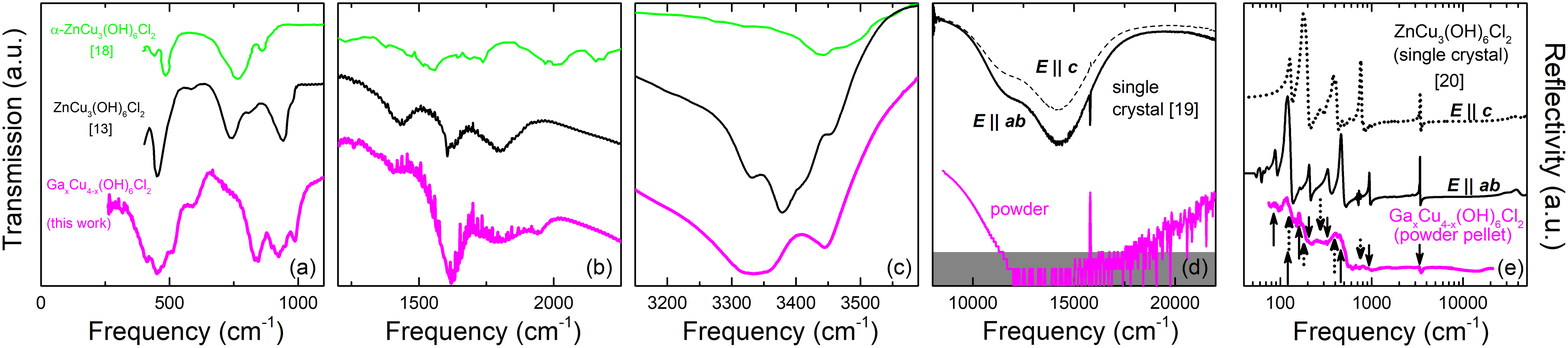}
\caption{(a-c) Comparison of room temperature powder transmission spectra of kapellasite~\cite{Krause2006}, herbertsmithite~\cite{Puphal2017} and \GA\ ($x=0.5$). (d) The green color is caused by an absorption in the red part of the visible range due to local $d$-$d$ transitions~\cite{Pustogow2017}. The grey area indicates the measurement sensitivity. (e) The vibration peaks in the infrared range agree best with \Herb, as illustrated in the reflectivity data by arrows (length is a rough measure of intensity) indicating the positions in the in-plane (solid black) and out-of-plane (dashed black) spectra of herbertsmithite~\cite{Dressel2018}. The spectra in (a-c,e) are vertically shifted.}
\label{optics-comparison}%
\end{figure*}

\section{Electron doping}
In order to experimentally determine the oxidation state of Ga and Cu we performed XAS measurements on two different compositions. The results are shown in Figure~\ref{XAS} together with spectra from adequate reference compounds taken under the same experimental conditions. The ionic state of the Ga and Cu ions can be identified by comparing the onset energy and shape of the XAS spectra with that observed for the reference compounds. Due to the well known chemical shifts in XAS, the onset of the K edge XAS spectra of Cu or Ga differs by several eV for the different oxidation states which therefore can be readily distinguished. This is nicely seen when comparing the spectra of the CuO (Cu$^{2+}$) with Cu$_{2}$O (Cu$^{1+}$) or GaN (Ga$^{2+}$) with Ga$_{2}$O$_{3}$ (Ga$^{3+}$). From the data in Figure~\ref{XAS} it is evident that copper and gallium are exclusively found in the form of Cu$^{2+}$ and Ga$^{3+}$; the presence of other oxidation states can be excluded. 

The main interest of this compound is to understand whether there is some additional free charge compared to herbertsmithite, since if we assume the simple oxidation states of ionic bonding we get Ga$^{\text{III}}$Cu$_{3}^{\text{II}}$(OH) $_{6}^{-\text{I}}$Cl$_{2}^{-\text{I}}$ resulting in a negative charge. Resistance measurements with a Keithley multimeter show that the compound has strong insulating behavior of several G$\Omega$, which is also reflected in its transparency and green color.

To obtain a more detailed view of the conduction properties and the band structure, optical spectroscopy experiments have been carried out. Overall, the spectrum reveals insulating properties suggesting a large band gap of several eV like in herbertsmithite \cite{Pustogow2017}; the complete absence of metallic band transport rules out any effect of doping, impurities  or disorder. The dips in the infrared transmission spectra shown in Figure~\ref{optics-comparison}~(a-c) correspond to phonon peaks, which appear at similar frequencies as for \Herb\ and related compounds~\cite{Krause2006,Sun2016,Puphal2017,Pustogow2017,Dressel2018}. While the vibration modes at 1000~\cm\ and below are related to motions of oxygen and heavier units, the higher-frequency features include the light hydrogen entities, in particular the intense 3300--3400~\cm\ mode~\cite{Dressel2018}. Note that optical measurements were performed on $^1$H samples ($0.5\leq x \leq 0.8$) rather than the deuterated analogue \GAD. We find close similarities to herbertsmithite --~our reflectance data in Figure~\ref{optics-comparison}~(e) reveal a one-to-one correspondence among most vibration modes with respect to the polarized spectra of \Herb\ \cite{Dressel2018}~-- providing evidence that the material under study is isostructural. The broader peaks of the \GA\ powder most likely result from the activation of additional vibration modes with different symmetry, or simply larger oscillator strength due to variation of ionicity Ga$^{+3}\leftrightarrow$ Cu$^{+2}$ giving rise to distinct dipole moments. In \Herb\ Zn$^{+2}$ and Cu$^{+2}$ have the same ionicity but affect vibrations in different ways.  
From Figure~\ref{optics-comparison}~(d) we see that the compound strongly absorbs at the red end of the visible range, resulting in the green color of \GA. At these energies local $d$-$d$ transitions with well-defined peaks were resolved in \Herb\ and \Y\ single crystals \cite{Puphal2017,Pustogow2017}; here powder measurements yield rather broad absorptions, with the peaks being beyond our measurement sensitivity (gray area). Hence, we attribute the feature observed here to local $d$-$d$ processes; however, scattering due to particles of size comparable to the wavelength is expected to contribute as well.

\begin{figure}[b]
\centering
\includegraphics[width=1\columnwidth]{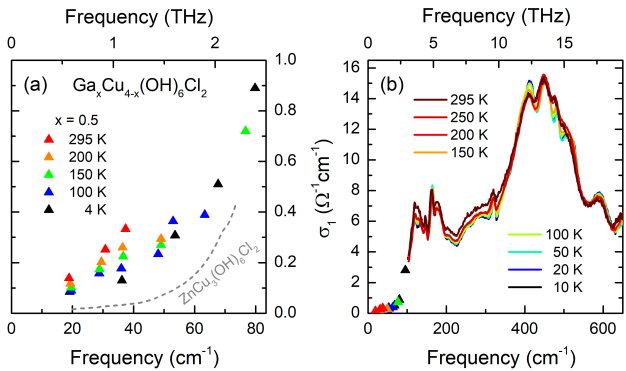}
\caption{(a) The temperature dependence of \GA\ ($x=0.5$) at the lowest frequencies shows insulating behavior with vanishing $\sigma_1(\omega\rightarrow 0)$ at all temperatures and freeze-out of charge excitations upon cooling. The optical conductivity is significantly larger than for herbertsmithite (here shown for $E\parallel ab$~\cite{Pilon2013}), but exhibits a similar power-law frequency dependence. (b) The optical conductivity in the far-infrared (solid lines) is dominated by phononic features on top of the insulating spectral response. Apart from modifications of the phonons around 400--600~\cm, the spectra and their temperature dependence are similar for higher substitutions up to $x=0.8$.}
\label{optics-temperature}%
\end{figure}

In order to identify possible spinon excitations in the electrodynamic response of the spin-liquid state~\cite{Ng2007,Balents2010,Dressel2018,Pustogow2018} we trace the low-energy absorption in the far-infrared and THz spectral ranges in Figure~\ref{optics-temperature}. The optical conductivity clearly vanishes towards $\omega\rightarrow 0$ in accord with our resistance measurements and there is only a weak temperature dependence below the phonon modes; this agrees with the rather large charge gap and the classification of the compound deep in the Mott insulating state analogue to herbertsmithite \cite{Dressel2018,Pustogow2017}. Below 100~\cm\ the optical conductivity is reduced upon cooling, indicating a typical freeze-out of thermal excitations as it was observed in several quantum-spin-liquid Mott insulators with a triangular lattice subject to strong electronic correlations~\cite{Pustogow2018Mott}. Our data give no indication for a non-thermal increase as reported for the in-plane component of \Herb\ \cite{Pilon2013}; however the background conductivity beneath the phonon peaks is significantly larger. The enhanced value and the insulating-like temperature dependence may be a result of measuring a mixture of all crystal directions in powder samples: the conductivity perpendicular to the kagome layers of herbertsmithite exhibits a conventional reduction upon cooling. Below 50~\cm\ we observe a power-law like tail similar to the one reported in \Herb\ \cite{Pilon2013}. Since this frequency dependence seems to persist up to room temperature, a relation to spin excitations is not straightforward -- 
experiments towards even lower energies, further away from the phonon modes and deeper within a possible spinon continuum~\cite{Pustogow2018} may clarify the situation.

We also measured \GA\ samples with larger substitutions $0.5\leq x \leq 0.8$, yielding very similar behavior in the entire frequency and temperature range discussed above. Some far-infrared vibrations show modifications as the Ga content increases, but the lowest-energy phonons at 120~\cm\ (involving the entire lattice and deforming the kagome layers) remain unchanged and very similar to herbertsmithite~\cite{Dressel2018}. 
After all, our optical experiments on compounds with different stoichiometry clearly rule out charge-carrier doping in \GA\ and place this material in the same category as herbertsmithite and related kagome systems, being a paramagnetic Mott insulator with a charge gap of several eV~\cite{Pustogow2017}.

\section{Magnetic susceptibility}

\begin{figure}[h]
\noindent \begin{centering}
\includegraphics[width=\columnwidth]{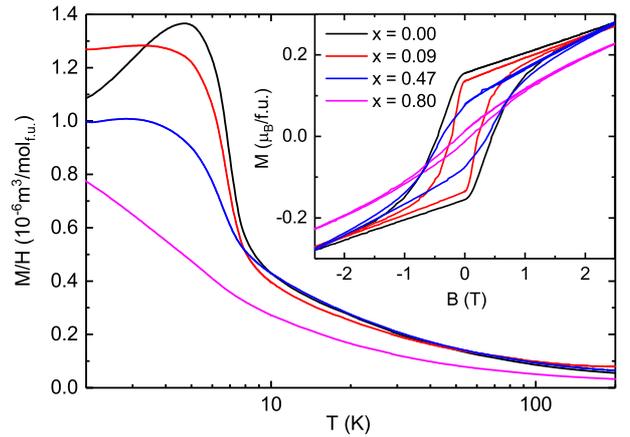}
\par\end{centering}
\caption{\label{M-Ga}Magnetization divided by magnetic field plotted against a logarithmic temperature scale for different Ga$_{x}$Cu$_{4-x}$(OD)$_{6}$Cl$_{2}$ powder samples measured at $\mu_{0}H=1$\,T. On the inset, the $M(H)$ curve measured at 2~K is shown revealing a hysteresis for all powder samples. }
\end{figure}

For clinoatacamite ($x=$\,0) a Curie-Weiss temperature of 190\,K was reported in Ref. \cite{Zheng2005}. A clear anomaly at 6.5~K could be observed both in specific heat as well as DC susceptibility associated to the long-range magnetic order of this compound. Furthermore, clinoatacamite  shows a phase transition at 18.1 K in specific heat, which is field dependent indicating a magnetic origin, although it can hardly be observed in the susceptibility. This transition is well resolved in $\mu$SR as will be discussed below. At 2\,K, a hysteresis with a remanent magnetization of 0.05\,$\mu_B$/Cu and a coercivity of 0.5\,T was measured \cite{Zheng2005}.  

In Figure~\ref{M-Ga} we present the susceptibility measurements of different Ga$_{x}$Cu$_{4-x}$(OD)$_{6}$Cl$_{2}$ batches. The overall behavior is quite similar to Zn$_{x}$Cu$_{4-x}$(OH)$_{6}$Cl$_{2}$.  For $x=0$ (black curve) a clear magnetic phase transition at $T_{N}=6.5$~K is observed. The absolute value of the DC susceptibility at low temperatures slowly decreases with increasing $x$, while $T_{N}$ is nearly unchanged. This behavior is confirmed in $M(H)$ measurements at 2~K shown on the inset of Figure~\ref{M-Ga}, where the same slow reduction of the frozen spin fraction is apparent. The remanence values are 0.1554; 0.1351; 0.0802; 0.0147 $\mu_{B}/\text{mol}_{\text{f.u.}}$ for the samples $x=$\,0; 0.09; 0.47; 0.80. A magnetic phase fraction estimate can be performed by dividing the substituted sample by the first value (x = 0): $\frac{M_{r}}{M_{r}^{0\text{\,M}}}\approx$ 87\%; 52\%; 9\%. The hysteresis loop areas of 4147; 1468; 2120; 440 $\mu_{B}T/\text{mol}_{\text{f.u.}}$ yields a rather small value for the x = 0.09 sample due to a high amount of CuO impurity phase. Yet we can estimate the magnetic phase fraction by the relative change $\frac{A}{A_{0\text{\,M}}}\approx$ 35\%; 51\%; 11\% in general good agreement with the remanence values. From a Curie-Weiss fit of the inverse susceptibility for temperatures above 220~K we observe an increase of the Weiss temperature with increasing substitution, as it was observed for Zn substitution \cite{Shores2005}. E.g. for $x=0.8$ we determined a Weiss temperature of $\Theta_{W}^{Ga}=-256$~K compared to $\Theta_{W}^{Zn}=-285$~K \cite{Shores2005}. 

\section{Specific heat}
\begin{figure}[h]
\noindent \begin{centering}
\includegraphics[width=\columnwidth]{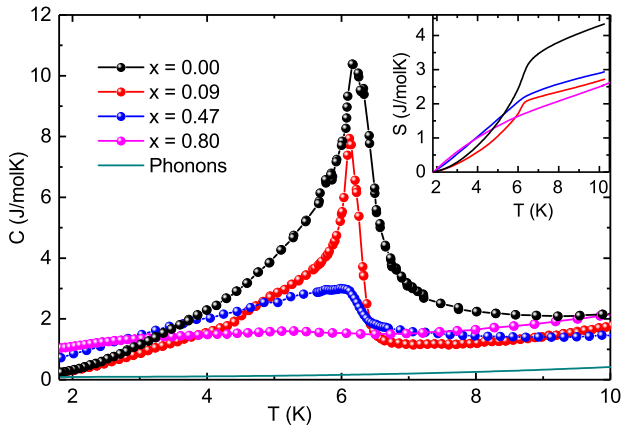}
\par\end{centering}
\caption{\label{spec heat Ga}Specific heat as function of temperature showing a $\lambda$-shaped peak revealing the magnetic order. The phonon contribution is not subtracted from this data, but an estimate shown by a turquoise line, which is much smaller than the magnetic contribution. As an inset the magnetic entropy $S$ is shown for the four different curves. }
\end{figure}

In Figure~\ref{spec heat Ga}, we present the specific heat on pressed pellets of Ga$_{x}$Cu$_{4-x}$(OD)$_{6}$Cl$_{2}$ for the same batches, where we have measured the DC magnetization. To overcome the weak thermal coupling in these samples, we mixed Apiezon N grease with powder and subtracted this contribution following Ref. \cite{Schnelle1999}. We used the same measurement procedure for all shown samples to get a good comparison. For clinoatacamite we reproduced the observed results of Ref.~\cite{Zheng2005}. With increasing amount of Ga we observe a pronounced decrease of this magnetic anomaly, but $T_{N}$ is only slightly shifted to lower temperatures with increasing $x$, in agreement to what was observed in the magnetization measurements (Figure~\ref{M-Ga}). In both measurements the strongest shift is observed in the $x=0.80$ sample with $T_{N}$ going down to about 5~K. The phonon contribution was estimated from the data of the $x=0.47$ sample, which were fitted in the range from 9\,K to 30\,K. The result of a linear fit of C$_{mol}/T$ vs. $T^{2}$ gives $\beta_{0}=0.329(4)$ mJ/(mol K$^{4}$), yielding a Debye temperature of $\vartheta_{D}\approx470$~K. We assume that this phonon contribution, $C_{ph}$, is similar for all different Ga batches and is plotted as a reference turquoise line in Figure~\ref{spec heat Ga} and is much smaller than the magnetic part for all measured samples.  

The entropy gain within the specific-heat anomaly at zero field is obtained by integration of $C/T$ and shown on the inset of Figure~\ref{spec heat Ga}. We calculated the contribution to the entropy of the magnetically ordered part by subtracting a linear background $S_{l}$ to disentangle dynamic spin fluctuations from the ordered part and obtain values of $S=\int_{1.8}^{10}C_{mol}/T\cdot dT-S_{l}\approx$\,2.82; 1.65; 0.43; 0.13\,J/molK$^{2}$. Thus we can estimate the magnetic phase fraction for the different concentrations to $S/S_{x=0}\approx$\,56\%; 15\%; 5\%. The variation in comparison to the susceptibility data is reasonable since we do not address the moments directly and the signal becomes quite broad and is going below the measured 1.8~K for increasing substitution.

\begin{figure}[t]
\noindent \begin{centering}
\includegraphics[width=\columnwidth]{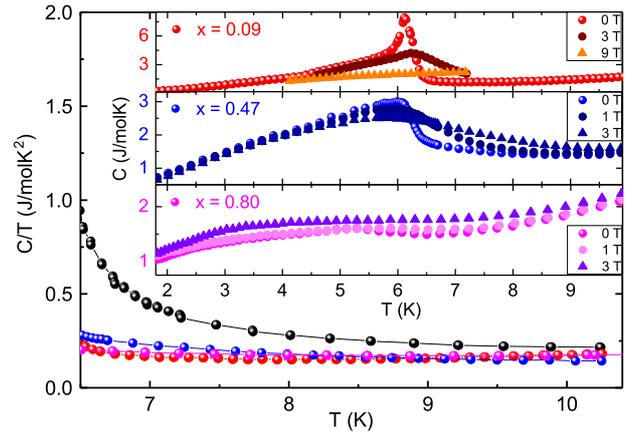}
\par\end{centering}
\caption{\label{spec heat field dep Ga} $C/T$ as a function of temperature above the magnetic order. Black dots show the measurements for $x=0$. As an inset the specific-heat measurements in different fields from 1.8 to 10\,K on the same exemplary batches show a slow washing out of the lambda shaped peak with increasing field.}
\end{figure}

A further qualitative approach to estimate the size of the ordered moment can be made by a comparison of the height of the anomaly. The values of the maximum are 10.4; 7.9; 3.0; 1.6\,J/(molK), respectively. Again the magnetic phase fraction can be estimated by dividing through the clinoatacamite value: $C/C_{x=0}\approx$ 72\%; 29 \%; 15\%. This shows that the ordered moment of the Ga$_{x}$Cu$_{4-x}$(OD)$_{6}$Cl$_{2}$ is slowly reduced and a large portion of the spin degrees of freedom remain fluctuating. Note that clinoatacamite is already a slightly frustrated system and does not order completely. Due to these enhanced spin fluctuations the specific heat follows a linear-in-$T$ dependence from $T_{N}$ up to 10\,K as can be seen in Figure~\ref{spec heat field dep Ga}. The field dependences of the specific-heat curves for the three concentrations show an overall similar behavior (inset of Figure~\ref{spec heat field dep Ga}). For $x=0.09$ the sharp peak is strongly suppressed with increasing field, leading to a broad maximum at 9~T. For $x=0.47$ a weaker broadening with field is observed. Only very small changes are visible for the $x=0.80$ sample, where the long-range magnetic order is already strongly suppressed at zero field.
 
\section{\label{subsec:SR-on-deuterated}Muon Spin Relaxation}
\begin{figure}[h]
\includegraphics[width=1\columnwidth]{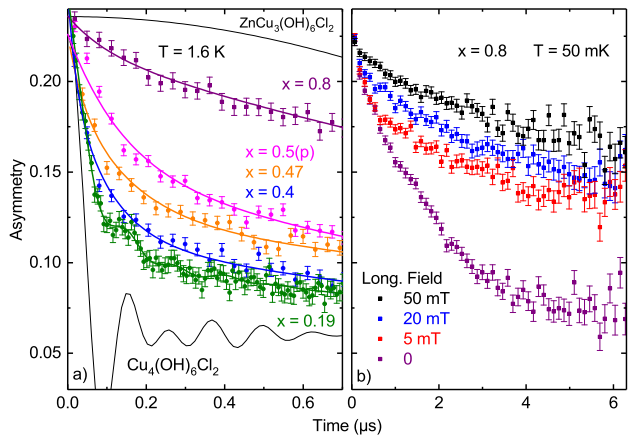} \caption{\label{musr_spectra}a) Muon decay asymmetry in zero-field at 1.6~K in Ga$_{x}$Cu$_{4-x}$(OD)$_{6}$Cl$_{2}$ with $x=0.19,\,0.4,\,0.47,\,0.8$ and a protonated version with $x=0.5(2)$ added to better span the phase diagram. Color lines are fits to a two component model accounting for a frozen and a paramagnetic volume fraction. The relaxation in herbertsmithite and clinoatacamite (thin solid lines) is reproduced from Ref.~\cite{Mendels2007,Zheng2005}. b) Relaxation at 50~mK in the $x=0.8$ sample in zero field and applied longitudinal fields. Note the different time scale as compared to a).} 
\end{figure}

In order to investigate locally the impact of the Ga substitution on the magnetic properties and compare with the Zn case~\cite{Mendels2007} we have performed muon spin relaxation experiments.  Figure~\ref{musr_spectra}a) shows the relaxation at 1.6~K of the muon asymmetry for 5 samples with increasing Ga concentration from $x=0.1$ to $x=0.8$. For comparison, the relaxations in herbertsmithite and clinoatacamite ($x=0$) are also reproduced from Ref.~\cite{Mendels2007,Zheng2005}. The relaxation in clinoatacamite ($x=0$) shows well resolved spontaneous oscillations as a result of well defined internal fields in its long range ordered ground state~\cite{Zheng2005}. Upon Ga substitution, these oscillations are rapidly smeared out, making evident the increasing magnetic disorder in the magnetic phase. Only in the least substituted $x=0.19$ sample weak and damped oscillations are still detected. At the same time a paramagnetic fraction develops giving rise to a slow exponential relaxation. In the $x=0.8$ sample this slow relaxation is largely dominant at $T=1.6$~K, thus resembling the purely dynamical behavior of the herbertsmithite spin liquid material. The relaxation was further measured down to 50~mK in this  sample and was found to remain mostly dynamical as demonstrated in Figure~\ref{musr_spectra}b) by the slow exponential shape and the weak field sensitivity~\cite{Schnelle1999}. Therefore a large fraction of the $x=0.8$ sample satisfies the first criterion for a spin liquid ground state, namely the absence of on site static magnetism down to $T/J\rightarrow0$.

\begin{figure}[h]
\includegraphics[width=\columnwidth]{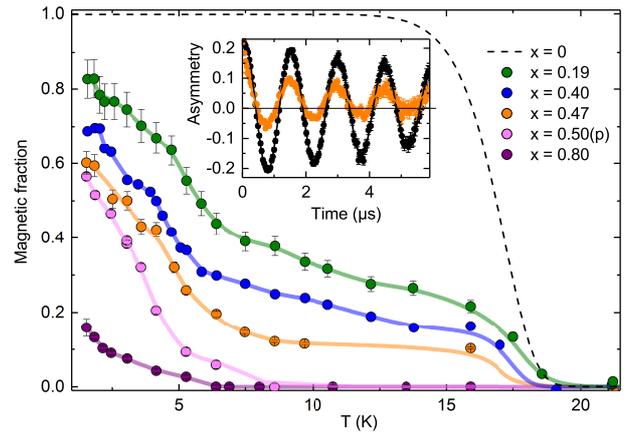} \caption{\label{magnetic_fraction}Temperature dependence of the volume fraction of the frozen magnetic phase for the same Ga doped samples as in Figure~\ref{musr_spectra}. For comparison, the magnetic fraction in clinoatacamite $x=0$ is reproduced from Ref.~\cite{Mendels2007} as a dashed line. Inset: example of the time-evolution of the muon decay asymmetry in the $x=0.47$ sample in a 5~mT transverse field at 21~K (black symbols) above the transition and at 1.6~K (orange symbols) in the ground state. Solid lines are fits to a damped cosine function of the long time ($t>1~\mu$s) asymmetry. The amplitude of the cosine is proportional to the non-magnetic fraction. The complementary magnetic fraction is plotted in the main panel.}
\end{figure}

\begin{figure}[h]
\noindent \begin{centering}
\includegraphics[width=\columnwidth]{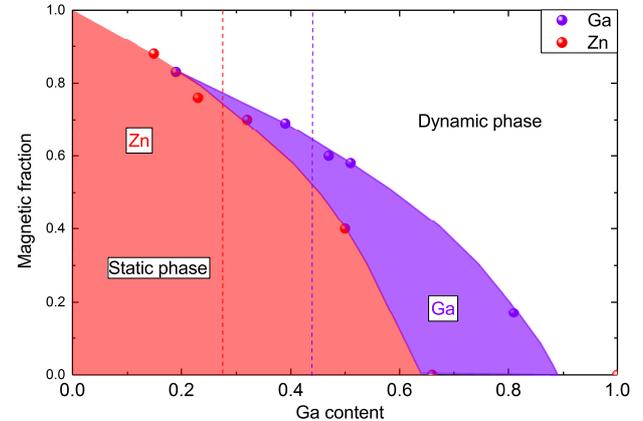}
\par\end{centering}
\caption{\label{phase diagram Ga}Phase diagram of the magnetic phase fraction versus the substitution content for Ga$_{x}$Cu$_{4-x}$(OD)$_{6}$Cl$_{2}$ (purple) compared to Zn substitution (red). A dashed vertical line marks the vanishing of a high-temperature order at 18\,K associated to the one observed in clinoatacamite \cite{Zheng2005}.}
\end{figure}

Additional measurements with a weak applied field $H_{TF}=5$~mT, transverse with respect to the initial muon polarization, were performed as a function of temperature to track the evolution of the paramagnetic and frozen volume fraction. Only muons stopping in the paramagnetic regions in the sample have their spin rotating freely at the frequency $\gamma_{\mu}H_{TF}/2\pi$ ($\gamma_{\mu}/2\pi=135.5$~MHz/T being the muon gyromagnetic ratio) whereas the strong internal fields in the frozen regions lead to a fast depolarization of the muon spin (see Figure~\ref{magnetic_fraction}). Thus, the amplitude of the oscillating component is proportional to the volume fraction of the sample which is not magnetically frozen. The substitution and temperature dependence of the complementary magnetic fraction is shown in the main panel of Figure~\ref{magnetic_fraction}. At low substitution, $0<x<0.5$, the transition occurs in two steps. A first transition is observed close to 18~K as in clinoatacamite but only a minority fraction of the sample becomes magnetic. Only below 6~K, does the magnetic fraction increase significantly to become dominant. These two subsequent transitions are strongly reminiscent of the two transitions observed in clinoatacamite but a noticeable difference is that the latter becomes fully magnetic already at 18~K and remains as such through the poorly understood 6~K one. For higher substitution $x\geq0.5$, the transition temperature is reduced to approach 6~K in the $x=0.8$ compound and the magnetic fraction is  strongly suppressed in favor of a dominant dynamical phase.
 
The magnetic behavior of the Ga-substituted atacamite as seen by $\mu$SR measurements is, overall, quite similar to the one of Zn-substituted atacamite. However in the Zn case~\cite{Mendels2007}, the two step transition is observed only at very low substitution $x=0.15$. One may speculate that the two step transition is observed only in weakly substituted compounds with the monoclinic space group of clinoatacamite while at stronger substitution, the symmetric rhombohedral phase of atacamite is  stabilized, corresponding to the disappearance of the 18~K magnetic transition. Within this assumption, the monoclinic to rhombohedric transition would be around $x=0.5$ in the Ga case and lower, in between $0.15$ and $0.33$, in the Zn one. Also remarkably, static magnetism is less efficiently suppressed by Ga substitution; the Zn$_{0.66}$Cu$_{3.33}$(OH)$_{6}$Cl$_{2}$ was found to be fully dynamical, spin-liquid like, in its ground state while in the $x=0.8$ Ga counterpart static magnetism is still detected as a $\sim20\%$ fraction. This is evident in Figure~\ref{phase diagram Ga} which compares the magnetic fractions for the two systems as a function of the dopant content. The transition temperature to three-dimensional long range order is likely set by the transverse interaction in between the kagome layers going through the Cu2-OD-Cu1 bonds. It thus seems that this interlayer coupling is slightly different in the Ga and Zn doped compound families.

\section{Nuclear Magnetic Resonance}
\begin{figure}[t]
\centering \includegraphics[width=\columnwidth]{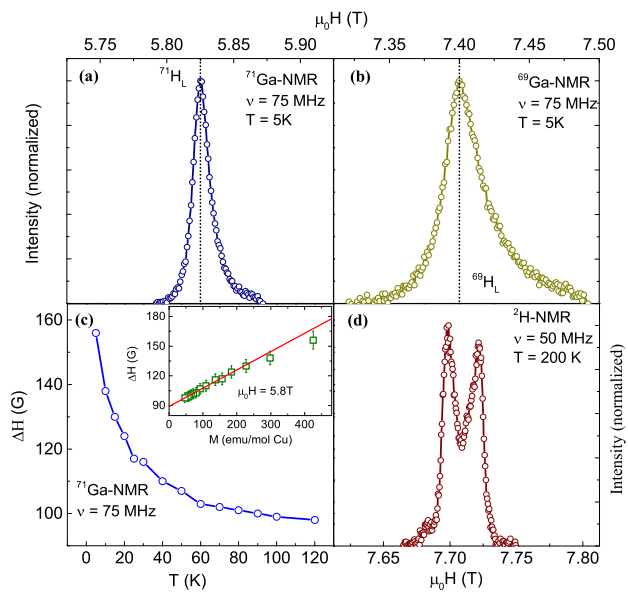}\caption{\label{nmr2} (a,b) $^{69,71}$Ga field sweep powder NMR spectra for the $x=0.8$ sample taken at 5~K. Vertical dashed lines indicate the Larmor field determined from a non-magnetic reference. (c) Temperature dependent line width of the $^{71}$Ga NMR line together with a plot of the width versus bulk susceptibility taken at the same field of 5.8 T (inset). (d) $^{2}$H deuterium NMR spectra taken at 200 K. }
\end{figure}

\begin{figure}[t]
\centering \includegraphics[width=\columnwidth]{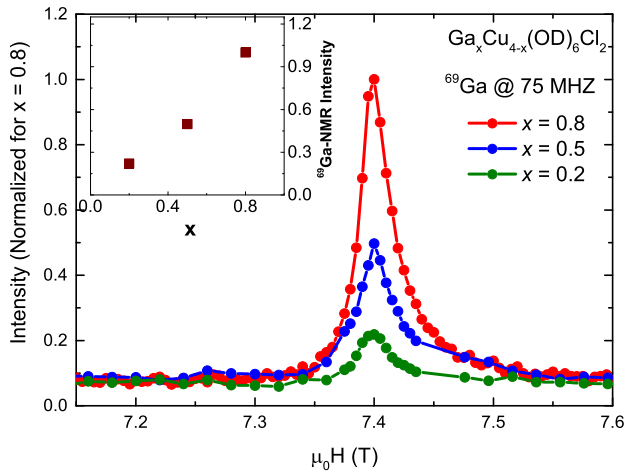} \caption{\label{nmr1} $^{69}$Ga NMR line at 5 K for three different Ga concentrations. The inset plots the integrated intensity versus Ga concentration.}
\end{figure}

For a deeper understanding of the local magnetic environment of the Ga-substituent, we have performed field sweep $^{69,71}$Ga NMR for the largest Ga concentration ($x=0.8$) at the NMR frequency of 75 MHz. The spectra for the $^{69}$Ga line are shown in Figure~\ref{nmr2}(a) and the $^{71}$Ga line in Figure~\ref{nmr2}(b). Both lines are at the expected resonance fields (with an ratio of $^{69}H/^{71}H$ = 1.270 ) and show a reasonable line broadening typical for Ga in Cu-based quantum magnets \cite{Khuntia2016}. The intensity ratio determined by the integration of $^{69}I/^{71}I = 1.27$ is close to the expected value of 1.5. From that we have strong evidence that the two lines are from the two Ga isotopes. There are no other NMR lines at higher and lower fields associated to either additionally sites or quadrupolar satellites. The satellites expected for an $I = 3/2$ line are hidden under the broadened central transition. We assume that the increase of the line broadening when lowering temperature is of magnetic origin. If so the width of the line should follow the bulk susceptibility \cite{Ranjith2018}. The $^{71}$Ga NMR line width (full width at half maximum, $\Delta H$) is plotted in Figure~\ref{nmr2}(c) as a function of temperature. The inset shows the width versus the molar magnetization $M$ taken in the same field of about 5.8~T. The linear behavior shows the direct relation and confirms the dominantly magnetic origin of $\Delta H$ at low temperatures. From $\Delta H$ vs $M$ we estimated an hyperfine coupling constant of 1.03~kOe/$\mu_{B}$ which is a reasonable number for Ga in Cu based quantum magnets \cite{Khuntia2016}. The fact that there is a residual $\Delta H_0$ term at high temperatures points towards a $T$-independent residual width, probably caused by $T$-independent quadrupole interaction. The $^{2}$H NMR spectra taken at 200~K at a frequency of 50~MHz is shown in Figure~\ref{nmr2}(d). The typical Pake-type spectra is rather narrow and indicates the absence of strong structural disorder at the deuterium site. Towards low temperatures this line gets also broadened very similar to the Ga NMR lines. We could not evidence any non monotonous broadening below 10~K due to spin freezing effect in $^{69,71}$Ga and $^{2}$H NMR. Nonetheless antiferromagnetic spin freezing effects are wiped out under the presence of strong magnetic fields, what was observed in specific heat and which is in contrast to zero field $\mu$SR where such effect become visible. From the NMR data taken at the $x=0.8$ sample we could state that we found well resolved NMR lines with reasonable broadening and they might be assigned to a single Ga site in the lattice.

Figure~\ref{nmr1} shows the $^{69}$Ga line at 75~MHz for different samples ($x =$ 0.19, 0.47, and 0.8) measured under the same conditions. Even for lower Ga concentrations there is only one $^{69}$Ga line detected. It becomes evident that the lines are not strongly broadened or shifted for different Ga concentrations. The absolute NMR intensity increases with increasing Ga content. Inset of Figure~\ref{nmr1} shows the NMR intensity as a function of Ga concentration $x$ which follows nearly a linear behavior. This clearly indicates a rather homogeneous Ga distribution in the samples.

\section{Discussion}{\label{SecDis}}
Since the insulating behavior disagrees with the theoretical proposal of a Dirac metal for $x=1$, GaCu$_{3}$(OH)$_{6}$Cl$_{2}$, there are different possibilities where the additional charge of the Ga is compensated. From the XAS measurements it is clear that Cu$^{1+}$ or Ga$^{1+}$ formation can be excluded. The charge could be balanced by vacancies on the Ga site. This scenario is supported by the fact that in most EDX images for $x>0.4$, the Ga content is lower than the stoichiometric ratio, but it is in strong contrast to the fact that we observe magnetic order in the $x=0.8$ sample, since in herbertsmithite at $x=0.8$ the order is already suppressed and with vacancies on the interlayer site less substitution would be necessary to expel Cu on this position between the kagome layers. Furthermore, in this case a much broader Ga NMR line would have been expected. A different possible scenario would be an additional Ga position between the kagome layers, where additional Ga-Ga bondings balance the charge. This would have led to different oxidation states for Ga, which would be visible in XAS spectra and can be also excluded. 

Therefore, the additional electron is most likely balanced by additional Cl or OH, but not in a stoichiometric manner, as observed in EuCu$_{3}$(OH)$_{6}$Cl$_{3}$\cite{Puphal2018} and YCu$_{3}$(OH)$_{6}$Cl$_{3}$ \cite{Puphal2017}, but more likely in a mixed form, e.g. as Ga$_{x}$Cu$_{4-x}$(OD)$_{6+2x/3}$Cl$_{2+x/3}$, which then is more difficult to detect in EDX and x-ray diffraction, because the determination of the oxygen is not as accurate as for Cl. But from the average over many EDX measurements we conclude that we have systematically an increased O and Cl content, as discussed in Sec.~\ref{Sec:EDX}.  

\begin{table}[h]
\caption{\label{tab:Summary-of-the}Summary of the frozen magnetic fraction for the 4 different Ga-concentrations, from the presented physical quantities. }
\noindent \begin{centering}
\begin{tabular}{|c|c|c|c|c|}
\hline 
batch & 0.0 M & 0.1 M & 0.7 M & 1.0 M\tabularnewline
\hline 
\hline 
x (EDX) & 0.00(0) & 0.09(6) & 0.47(8) & 0.8(1)\tabularnewline
\hline 
\hline 
$M_{r}\,[\mu_{B}/\text{mol}]$ & 0.1554 & 0.1351 & 0.0802 & 0.0147\tabularnewline
\textcolor{blue}{$M_{r}/M_{r}^{0\text{\,M}}$ {[}\%{]}} & \textcolor{blue}{100} & \textcolor{blue}{87} & \textcolor{blue}{52} & \textcolor{blue}{9}\tabularnewline
\hline 
\textcolor{black}{$A_{\chi}\,[\mu_{B}K/\text{mol}]$} & \textcolor{black}{4147} & \textcolor{black}{1468} & \textcolor{black}{2120} & \textcolor{black}{440}\tabularnewline
\textcolor{blue}{$A/A_{0\text{\,M}}$ {[}\%{]}} & \textcolor{blue}{100} & \textcolor{blue}{35} & \textcolor{blue}{15} & \textcolor{blue}{11}\tabularnewline
\hline 
\textcolor{black}{$S_{order\,}[\text{J/molK}^{2}]$} & \textcolor{black}{2.82} & \textcolor{black}{1.65} & \textcolor{black}{0.43} & \textcolor{black}{0.13}\tabularnewline
\textcolor{blue}{$S/S_{0\,\text{M}}$ {[}\%{]}} & \textcolor{blue}{100} & \textcolor{blue}{56} & \textcolor{blue}{15} & \textcolor{blue}{5}\tabularnewline
\hline 
\textcolor{black}{$C_{max}$ {[}J/(molK){]}} & \textcolor{black}{10.4} & \textcolor{black}{7.9} & \textcolor{black}{3.0} & \textcolor{black}{1.6}\tabularnewline
\textcolor{blue}{$C/C_{0\,\text{M}}$ {[}\%{]}} & \textcolor{blue}{100} & \textcolor{blue}{72} & \textcolor{blue}{29} & \textcolor{blue}{15}\tabularnewline
\hline 
\textcolor{blue}{Magnetic fraction} &  & $x=0.19$ &  & \tabularnewline
\textcolor{blue}{$\mu$SR {[}\%{]} at 1.6 K} & \textcolor{blue}{100} & \textcolor{blue}{82} & \textcolor{blue}{60} & \textcolor{blue}{17}\tabularnewline
\hline 
\end{tabular}
\par\end{centering}
\end{table}

Concerning the magnetic behavior, all measurements show consistently, that the amount of frozen spins slowly reduces with increasing Ga substitution. A comparison of the remanence $M_{r}$, area of hysteresis $A_{\chi}$, entropy $S_{order}$ and maximum of specific heat $C_{max}$ and their resulting frozen spin fractions is given in Table~\ref{tab:Summary-of-the}. The index 0~M of these quantities denotes the value of the 0.0 M batch.  The $\mu$SR measurements give the most direct access to the magnetic fraction and thus should be taken more serious than the other values, but in general, all different quantities show a continuous suppression of long-range magnetic order, similar to what was observed for Zn$_{x}$Cu$_{4-x}$(OD)$_{6}$Cl$_{2}$.  

\section{Conclusion}
In conclusion, Ga$_{x}$Cu$_{4-x}$(OD)$_{6}$Cl$_{2}$ has no Dirac metallic properties for $0\leq x\leq0.8$. In contrast, it is a frustrated quantum spin system with magnetic properties similar to Zn$_{x}$Cu$_{4-x}$(OD)$_{6}$Cl$_{2}$. The phase formation was successful by the reflux synthesis only at low temperatures, yielding powder samples with varying Ga content. The crystal growth through a hydrothermal technique, as it was successful for herbertsmithite, is therefore not applicable for Ga$_{x}$Cu$_{4-x}$(OD)$_{6}$Cl$_{2}$. A homogeneous Ga distribution through the samples was proven by rather narrow Ga NMR lines, whose line width did not significantly increase with higher Ga content. From optical spectroscopy, it is evident that for $x\leq0.8$ the obtained green powder is a Mott insulator. The extra electron seems to be compensated by additional OH$^{-}$/ Cl$^{-}$ anions. Therefore, electron doping of a two-dimensional kagome lattice remains a challenge for the future. Magnetization, specific heat and $\mu$SR measurements show consistently that the magnetically ordered fraction is continuously suppressed with increasing Ga substitution, but the general effect is slightly weaker compared to Zn substitution, revealing a finite ordered fraction for the $x=0.8$ compound. The determination of the spin-liquid properties in Ga$_{0.8}$Cu$_{3.2}$(OD)$_{6}$Cl$_{2}$ and their comparison to the observations in the Zn-counterpart remains a task for future work. 

\begin{acknowledgement}
The authors acknowledge support by the German Science Foundation (DFG) through grant SFB/TR49, DR228/39 and DR228/51 and by the French Agence Nationale de la Recherche under Grant ANR-12-BS04-0021 'SPINLIQ'. We further acknowledge stimulating discussions with R. Valent\'i.
\end{acknowledgement}

\section*{\small{Author contributions }}
\small{$ $ P.P. and M.M. synthesized the samples. P.P. measured EDX, PXRD, magnetic and specific heat under supervision of C.K. NMR measurements have been performed and analyzed by K.M.R. and M.B. Optical characterization was done by A.P. and M.D. XAS data were collected by A.R. and K.K. and $\mu$SR Data were measured and analyzed by E.K., F.B., and P.M. The work was initiated by P.P. and C.K. } 

\providecommand{\WileyBibTextsc}{}
\let\textsc\WileyBibTextsc
\providecommand{\othercit}{}
\providecommand{\jr}[1]{#1}
\providecommand{\etal}{~et~al.}

\end{document}